\documentclass[aps,prc,twocolumn,superscriptaddress,preprintnumbers,showpacs,floatfix,nofootinbib]{revtex4-1}
\usepackage{amsmath,graphicx,color,ulem}
\usepackage{hyperref,cleveref}
\newcommand{\trento}{T$\mathrel{\protect\raisebox{-2.1pt}{R}}$ENTo }

\begin{document}

\title{Signals for fluctuating constituent numbers in small systems}

\author{Andreas Kirchner}
\email{andreas.kirchner@duke.edu}
\affiliation{Department of Physics, Duke University, Durham, NC 27708, USA}

\author{Steffen A. Bass}
\email{bass@duke.edu}
\affiliation{Department of Physics, Duke University, Durham, NC 27708, USA}

\date{\today}

\begin{abstract}
We propose an extension of the  initial condition model T$_\text{R}$ENTo for sampling the number of partons inside the nucleons that participate in a heavy-ion collision. This sampling method is based on parton distribution functions (PDFs) and therefore has a natural dependence on the momentum transferred in the collision and the scale being probed during the collision. We examine the resulting distributions and their dependence on the momentum transfer. Additionally, we explore the sensitivity of different observables on the number of partons using the T$_\text{R}$ENTo framework and the estimators available therein for final-state observables.
\end{abstract}
\maketitle

\section{Introduction}

One of the major tools for enhancing our understanding of quantum chromo dynamics (QCD) are ultra-relativistic heavy-ion collisions. These collisions, conducted mainly at the large hadron collider (LHC) at CERN and the relativistic heavy-ion collider (RHIC) at Brookhaven National Laboratory \cite{Harris:1996zx,PHENIX:2003iij,PHENIX:2003qra,STAR:2008med,ALICE:2013mez,ALICE:2022wpn,Harris:2023tti}, allow to probe the properties of hot and dense QCD matter by comparison between experimental data and theoretical models \cite{Bernhard:2016tnd,Bernhard:2019bmu,Nijs:2020roc,JETSCAPE:2020mzn}. A majority of these models use a fluid-dynamic description for the bulk part of the collision and encode the underlying QCD dynamics through the equation of state and transport coefficients, such as shear and bulk viscosities. While it is possible to obtain the equation of state and the transport coefficients through means such as lattice QCD, renormalization group equations and others \cite{HotQCD:2014kol,Christiansen:2014ypa,Policastro:2001yc}, the description of the first few moments of the collision from first principles is much harder. Despite recent advances in this area \cite{Schenke:2012wb,Niemi:2015qia,Kurkela:2018vqr}, the precise dynamics is not yet fully understood and computationally expensive to study. A popular alternative choice is the usage of initial state models based on more parametric ansatzes, such as Monte Carlo Glauber models \cite{Miller:2007ri,Loizides:2017ack} or the T$_\text{R}$ENTo initial condition model \cite{Moreland:2014oya,Ke:2016jrd,Moreland:2018gsh,Soeder:2023vdn}. Since these models include parameters that may be difficult to directly relate to calculable QCD quantities, a common way of determining their value is by calibrating the respective models to experimental data. Depending on the type of measurement at hand, a different set of parameters can be constrained based on the model-data comparison.

One recent example regarding the discovery of new parameters for calibration was the search for the chiral magnetic event at RHIC \cite{STAR:2021mii} through the collision of two isobar isotopes ($^{96}_{44}\mathrm{Ru}$ and $^{96}_{40}\mathrm{Zr}$). Extensive studies have demonstrated a strong dependence of the ratios between isobars on nuclear structure parameters, such as the neutron skin thickness and deformations \cite{Giacalone:2025vxa,Xu:2021uar,Xu:2021vpn,Jia:2021oyt,Nijs:2021kvn,Li:2018oec}.

While the the modeling of deformed nuclei is well established and the relation between the nuclear deformation and their effects on flow coefficients in particular has been demonstrated \cite{Giacalone:2024luz}, the modeling of sub-nucleonic degrees of freedom for hydrodynamic initial conditions remains an active area of research in the heavy-ion community \cite{Moreland:2018gsh,Nijs:2020roc,Soeder:2023vdn,Shen:2017bsr,Giacalone:2021clp}. A common way for including sub-nucleonic structure in the \trento model is the addition of $m$ partons inside each proton and neutron with $m$ ranging from three to values of around ten, with $m$ being a fixed number for all nucleons in both ions. However, from deep inelastic scattering and other experiments, it is known that protons and neutrons have a rich substructure, especially when probing them at small Bjorken $x$, also known as saturation region \cite{Kovchegov:2014kua}. This dynamics is encoded in so-called parton distribution functions \cite{Lai:1999wy}, which give the probability to find a parton at a given Bjorken $x$. Additionally, the number of partons for any nucleus at a given momentum transfer is not fixed, but fluctuates on an event-by-event basis. In this paper we present a novel way of sampling the number of constituents of a nucleon in \trento, based on parton distribution functions, accounting for event-by-event fluctuations of the number of partons.

The sampling algorithm together with the resulting distributions are discussed in Sec. \ref{sec_model}. The signature of these fluctuations is then examined and shown to be strongest in small collision systems, such as oxygen or neon, (Sec. \ref{sec_sensitivity}). Additionally, we examine different observables and determine a combination of known observables displaying an enhanced sensitivity to the parton number in Sec. \ref{sec_sensitivity}.

\section{Parton number sampling algorithm} \label{sec_model}

Before discussing the details of the sampling algorithm, we will first discuss how sub-nucleonic structure is currently treated in \trento:
In \trento the positions of the nucleons inside the nuclei are sampled based on the nuclear density. The probability for a collision is then determined based on the cross section and distance for each nucleon pair in the two colliding nuclei. Since this collision process is so far not well understood from first principles, the amount of energy deposited in each nucleon-nucleon collision is modeled by placing a Gaussian at the collision site, whose normalization is being drawn from a Gamma distribution to allow for fluctuations. At this step the \trento model allows for the inclusion of subnucleonic degrees of freedom, i.e. quarks and gluons, by means of depositing not one large Gaussian at the collision site, but rather depositing $m$ smaller Gaussians around it \footnote{The sampling for their positions is designed to yield the same nucleon size as the single Gaussian in the limit of large $m$}. It was already demonstrated that this fixed number of sub-nucleonic degrees of freedom can be constrained using bayesian inference with $m$ found to be between four and five\footnote{The number of constituents is to be understood as rounded value of the reported maximum likelyhood value of the posterior.} with rather large uncertainties \cite{Nijs:2021clz, Nijs:2020ors, Nijs:2020roc}. Additional, stronger constraints can arise in smaller collision systems \cite{Huang:2025cjm}. 



This model of a fixed number of hotspots at each collision site only superficially reflects our understanding of the structure of the proton and neutron for two main reasons:

Firstly, it is well established that protons and neutrons have complicated and rich substructure consisting out of three valence quarks together with a fluctuating amount of sea quarks and gluons, resulting in a fluctuating parton number when probing the same proton or neutron multiple times. The second discrepancy between the model and our understanding is the fact that if the proton had a fixed number of partons, not all would need to interact in the inelastic collision due to the different energy (and therefore time) scales set by the inelastic scattering compared to the scales in parton-parton interactions. Both of these effects lead to a fluctuating number of partons in each nucleon-nucleon collision.

To determine this fluctuating number of partons, we propose a novel distribution based on the parton distribution functions (PDFs) describing the momentum distribution of partons inside of the nucleus. The parton distribution functions hereby serve as probability distributions for finding a parton of a certain type in the momentum range from $x$ until $x +\mathrm{d}x$, with $x$ being the momentum fraction carried by the parton of the total momentum of the proton or neutron. Note that this is not a probability distribution function in the strict sense, since it is not normalized to unitiy for the different species. Nevertheless, we will use it to obtain the parton number for a given event based on the moment fractions sampled from the PDFs:

For the sampling of the parton number, we will assume that the three valence quarks found in the proton and neutron are strongly correlated among each other and therefore always participate in the reaction. Therefore, the sampling will begin by drawing the momenta of the three valence quarks from their respective up and down quark distribution functions. The sampled momenta are then added up, $\sum_{val} x_{val}=x_{tot}$. Afterward, depending on the momentum transfer $Q^2$ the species for the next sample is determined randomly from gluons, up, down, strange, charm and bottom quarks, where the heavier species are only sampled for large enough $Q^2$. Subsequently, either the momentum of one gluon or a quark anti-quark pair is sampled from the respective distribution. The drawn momenta are then added to the total momentum sum $x_{tot}$. This procedure beginning from the random species determination is then repeated until the total momentum sum exceeds one: $x_{tot}>1$. The total number of partons is then given by three for the valence quarks plus the amount of drawn sea gluons, quarks and anti-quarks. At this point it is important to note that the number of partons will be bounded from below, despite PDFs sharply rising for $x\to 0$, due to a lower cut-off imposed on $x$ by kinematic considerations:

Using the relation between the momentum transfer $Q^2$, the center of mass energy $s$, the target parton momentum fraction $x$ and the projectile energy loss $y$

\begin{equation}
    Q^2=sxy
\end{equation}

and using $y\leq 1$, we find the minimal value for $x$, given by

\begin{equation}
    x \geq \frac{Q^2}{s}=x_{min}.
\end{equation}

\begin{figure}[h]
\includegraphics[width=0.9\linewidth]{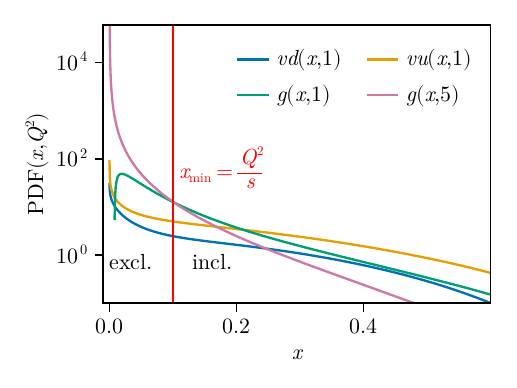}
\caption{Parton distribution functions as function of Bjorken $x$, taken from UCL HEP group \cite{Harland-Lang:2014zoa}. The distribution functions sharply rise for $x\to 0$, resulting in smaller values of $x$ being favored. The cutoff $x_{min}$ ensures that the PDFs are finite and have a maximum value, ensuring a finite number of partons in the sampling process.
}
\label{fig:PDFCutoff}
\end{figure}

In spite of the small value of $x_{min} \approx 10^{-6}$ for top LHC energies, it nevertheless provides a cut-off for the parton distribution function, ensuring a finite maximal value, see Fig.~\ref{fig:PDFCutoff}. While the parton distribution functions sharply rise for $x \to 0$, the cutoff $x_{min}$ ensures a finite maximal values for each distribution, resulting in a normalizable distribution. Furthermore, since the cut-off point is larger than zero $x_{min}>0$, it ensures a maximal value for the number of partons given by $m_{max} = 1/x_{min}$. The parton distribution functions used for the sampling are taken from the UCL HEP group \cite{Harland-Lang:2014zoa} \footnote{We want to highlight that ideally, the partons would be sampled using the $n$-parton distribution function (instead of the 1-parton distributions), including all correlations between the partons. However, since these are not yet known, we will sample partons individually, neglecting some of their correlations.}. 

\begin{figure}[h]
\includegraphics[width=0.9\linewidth]{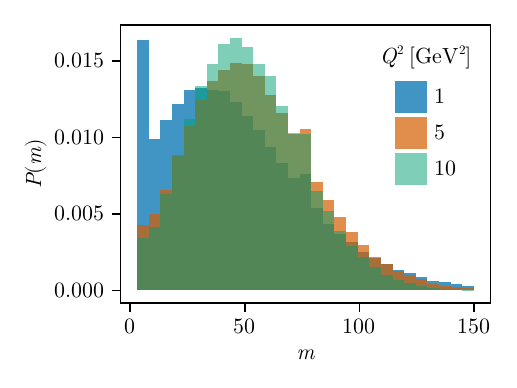}
\caption{Probability distribution for the number of partons for different values of $Q^2$. All distributions are bounded from below by three through the presence of the valence quarks.}
\label{fig:DifferentQDistr}
\end{figure}

The resulting distribution for the parton number $m$ for different values of $Q^2$ are displayed in Fig.~\ref{fig:DifferentQDistr}. All distributions are bounded from below by $m=3$ through the presence of high momenta valence quarks. Since the PDFs are less peaked for lower values of $Q^2$, there are more high momenta valence quarks at low $Q^2$, resulting in a distinct peak of the parton distribution at $m=3$. Since a collision with a larger momentum transfer probes the system at smaller length scales, the number of parton grows and therefore the maximum of the parton distribution also shifts to larger values.

\begin{figure}[h]
\includegraphics[width=1\linewidth]{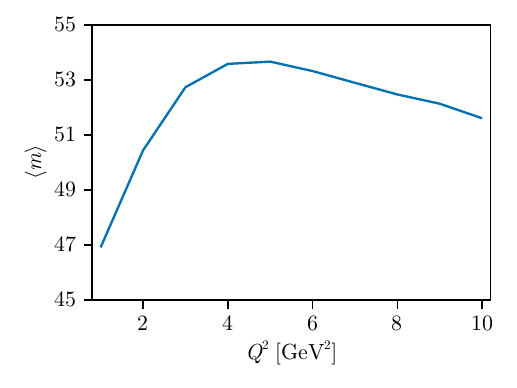}
\caption{Average number of partons in each nucleon as function of momentum transfer $Q^2$. The average number of partons increases with $Q^2$, since the parton distribution functions become more peaked towards $x=0$, due to the higher scattering resolution. The increase stagnates due to the increasing cutoff $x_{min}$.}
\label{fig:AveM}
\end{figure}

\begin{figure}[h]
\includegraphics[width=0.9\linewidth]{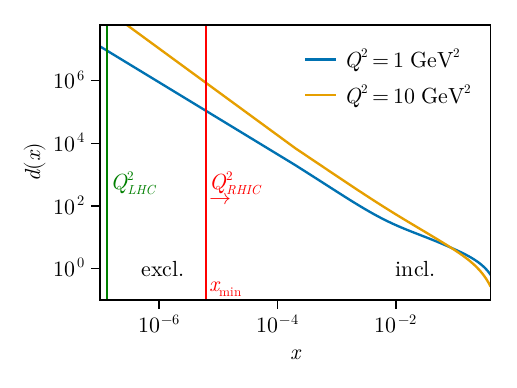}
\caption{Down quark PDF for different values of $Q^2$. The distribution function is more narrow for larger values of $Q^2$, driving the sampled $x$ towards $x_\text{min}$. Through the increase of this cut-off, the sampling range is restricted, leading to fewer fluctuations in the total number of partons.}
\label{fig:xmin}
\end{figure}

Similarly to the maximum of the distribution, also the average parton number grows with $Q^2$ initially, see Fig. \ref{fig:AveM}. However, for large enough $Q^2$, the average number of partons starts to decrease again. This can be attributed to the interplay between the scale being resolved in the scattering process getting smaller with larger $Q^2$ and the momentum cutoff getting larger with $Q^2$. The resolution of smaller length scales leads to the PDFs being more peaked toward $x\to 0$, resulting in more low momentum partons being sampled, increasing the total number of partons. However, with rising $Q^2$, the excluded momentum range below $x_{min}$ also grows, leading effectively to partons with higher momenta, therefore reducing the parton number (see Fig. \ref{fig:xmin}).


\begin{figure}[h]
\includegraphics[width=1\linewidth]{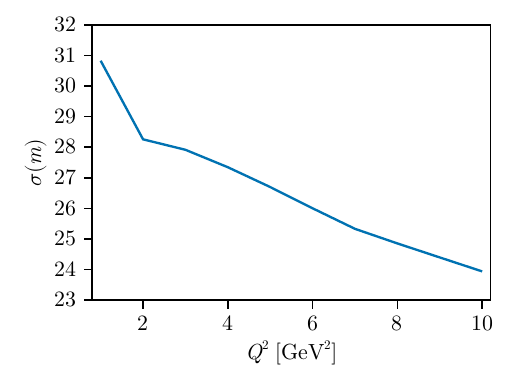}
\caption{Standard deviation of the parton number distribution as function of $Q^2$. The standard deviation decreases for increasing $Q^2$ due to the underlying distributions being more peaked at higher momentum transfer and a higher momentum cutoff $x_{min}$.}
\label{fig:stdM}
\end{figure}

This interplay between the changing distribution functions and the cutoff with changing momentum transfer also leads to a decreasing standard deviation of the parton distributions with increasing $Q^2$, as shown in Fig. \ref{fig:stdM}. The higher cutoff reduces the range the sampled momentum is able to fluctuate in, leading to a smaller standard deviation of the distribution of partons.

Since this method of obtaining the parton number solely relies on parton distribution functions, it is possible to employ this method for different collision systems and energies through the use of DGLAP equations \cite{Gribov:1972ri,Dokshitzer:1977sg,Altarelli:1977zs}. An additional extension using the different distributions for protons and neutrons is straight forward by using the respective proton or neutron distribution functions. In a similar fashion, the difference between different nuclei can be accounted for by using the nuclear distribution function instead of the proton or neutron ones.

Since the parton number can get large, especially for large center of mass energies, a large number of sampling processes can be required, resulting in a higher evaluation cost of initial state model, especially for heavy nuclei. A possibility to circumvent this added cost to the evaluation of the initial conditions is to employ a parameterization or fit of the parton distributions presented in Fig.~\ref{fig:DifferentQDistr}. A crude parametrization of the distribution can be obtained from its cumulants, with the first two (mean and standard deviation) being given in Fig.~\ref{fig:AveM} and Fig.~\ref{fig:stdM} as function of $Q^2$.

Additionally, the question arises, if one should take all of the sampled partons into account. Despite the large correlations between the individual partons in the nucleus, not all of them might participate in the interaction due to possibly very large virtualities. In the entirety of this work, we assume the correlations to be strong enough for all particles, no matter their virtuality, to interact. However, a further examination of such a "cutoff" in virtuality presents an intriguing possibility for the extension of this work.

Another extension of this sampling method is the usage of so-called transverse momentum distribution functions (TMDs) instead of parton distribution functions. Transverse momentum distribution functions can be understood as extensions of parton distribution functions by including additional information about the transverse momentum distribution of the individual partons. This additional information can be used to sample the transverse momenta of the partons participating in the collision. The inclusion of this information can then allow for the calculation of a initial non-zero fluid velocity.

While we believe that this model has a strong physics motivation and propose its implementation in the next version of an initial state model, such as \trento \footnote{The actual implementation requires additional considerations regarding the amount of deposited energy per parton depending on their respective momentum fraction.}, we will use calculations at fixed $m$ to examine the sensitivity of different observables on the number of partons using initial state estimators.


\section{Sensitivity of different observables} \label{sec_sensitivity}

In this section we want to gauge the sensitivity of different observables and collision systems on the proposed fluctuating number of constituents. In principle, every combination of observables can be used to constrain the number or distribution of partons inside each nucleon. However, some of these observables will show greater sensitivity than others. To examine these effects, we will make use of the well known proportionalities \cite{Giacalone:2020ymy} linking initial state quantities, such as the entropy, energy and eccentricities to the number of charged particles, the mean transverse momentum and the harmonic flow coefficients,
\begin{align}
    N_\text{ch} &\propto S, \\
    \langle p_\mathrm{T} \rangle &\propto E, \\
    v_n &\propto \epsilon_n.
\end{align}
Even though these are not precise relations, they nevertheless can aid in gauging the sensitivity to different observables by employing only initial state simulations, which are relatively cheap compared to a full fluid dynamical simulation. 

We already can establish tendencies of the sensitivity of different observables: Since the amount of energy carried by each nucleus is the same, irrespective of the number of partons, the energy deposited in each collision will remain similar. Likewise, the entropy of the collision will remain largely unchanged, although one could make the argument that the strong correlations between the individual partons will increase the entropy of the initial profile. Since the description of this entropy from first principles is non-trivial, we will assume that it can be neglected in this context. Having multiple constituents inside each nucleon might lead to more defined geometric features of the initial profile compared to the case of larger, smooth nuclei, resulting in a change of the initial eccentricities. Therefore, the largest signal is expected in the eccentricities.

\begin{figure}[h]
\includegraphics[width=1.0\linewidth]{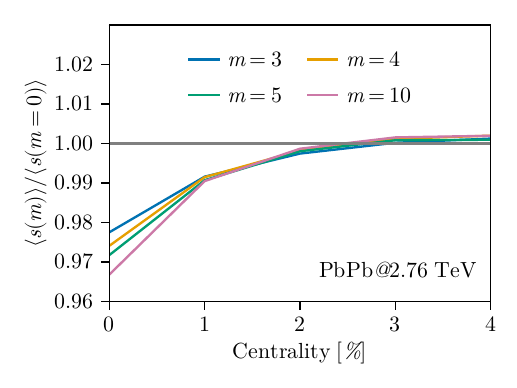}
\caption{Average entropy as function of centrality for different number of partons for central collisions. The most central collisions are sensitive to the number of partons through saturation of binary collisions and therefore can probe the average value of the distribution function for the parton number.}
\label{fig:EntroZoom}
\end{figure}

\begin{figure}[h]
\includegraphics[width=1.0\linewidth]{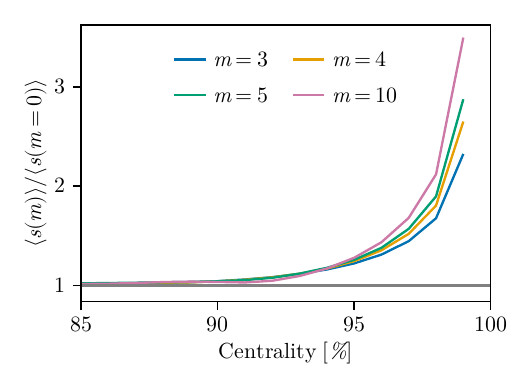}
\caption{Average entropy as function of centrality for different number of partons for ultra peripheral collisions. The ultra peripheral collisions only include a few binary collisions and therefore correspond to individual samples of the parton number distribution.}
\label{fig:EntroZoomHighCentra}
\end{figure}

Depending on the centrality and the collision system, up to 2000 binary nucleon-nucleon collisions take place in a single heavy ion collision, effectively sampling the distribution of constituents for each of the binary collisions. Regrettably, it is not possible to access these 2000 samples individually, which would allow to put precise constraints on the distribution and its moments. Since all of the individual energy and entropy deposition sites are combined into one initial energy/entropy density profile, even the initial condition used for fluid-dynamic simulations only contains an average of the partonic distribution function. More of this information will be washed out during the hydrodynamic evolution phase and subsequent particle production. Since this effective averaging scales with the number of binary collisions, we expect more insights into the parton distribution when considering events with a lower amount of binary collisions. This can either be achieved by considering more peripheral events (see Fig.~\ref{fig:EntroZoomHighCentra}) or by examining smaller collision systems. Interestingly, also ultra-central events show some dependence of the number of partons, even if only a slight one, as depicted in Fig.~\ref{fig:EntroZoom}. This stems from the number of spectators going to zero, resulting in a similar value of $N_\text{coll}$ for each event equaling no $N_\text{coll}$ fluctuations. When using a fluctuating distribution for the number of partons, this would result in probing the average of the distribution.


\begin{figure*}[h]
\includegraphics[width=0.75\linewidth]{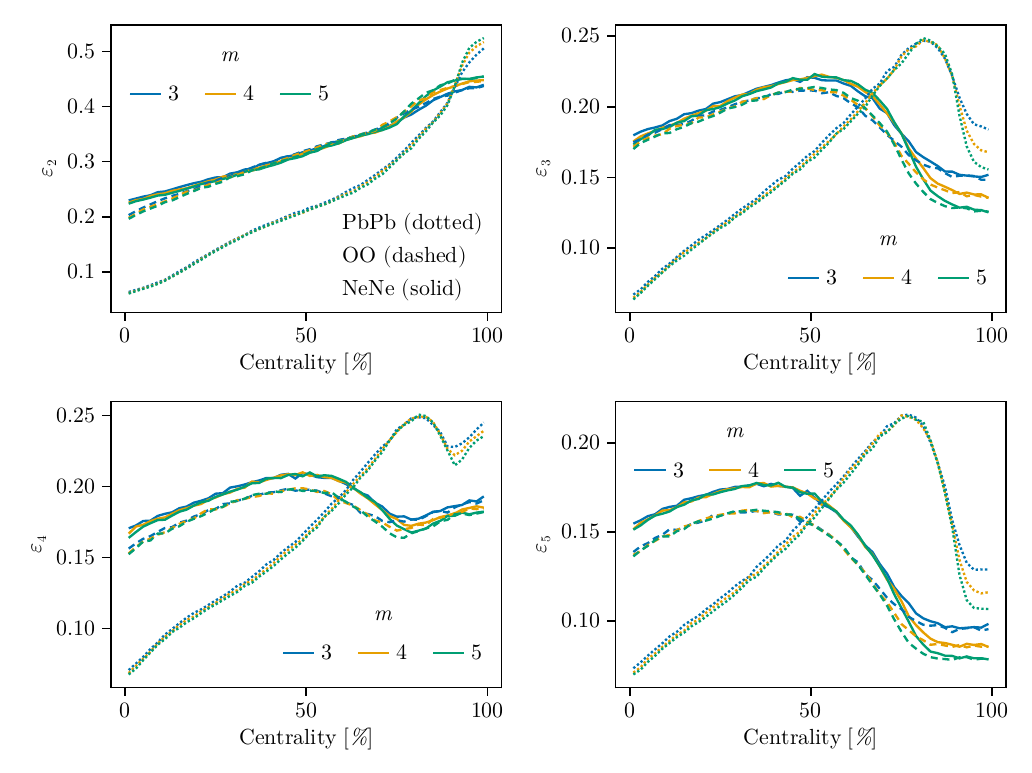}
\caption{Mean values of the first four harmonics as function of centrality for different collision systems and fixed values of $m$. The differences in sub-nucleonic structure appear around the same number of binary collisions, resulting in a larger centrality for PbPb collisions compared to OO and NeNe collisions. The even harmonics show less dependence on the value of $m$ compared to the odd harmonics, since they are mainly driven by the geometric shape of the initial entropy profile.}
\label{fig:FlowCoeffComp}
\end{figure*}

The expected difference in the eccentricities is demonstrated in Fig.~\ref{fig:FlowCoeffComp}, where the eccentricities as function of centrality are displayed for three different systems: PbPb, OO and NeNe collisions. Similarly to the initial entropy small deviations for the very central collisions also appear in the eccentricities. These differences are then much more pronounced for peripheral collisions. Here it is interesting to note that the even harmonics, i.e. $\varepsilon_2$ and $\varepsilon_4$, show less separation, especially for the large centralities, compared to the odd harmonics $\varepsilon_3$ and $\varepsilon_5$. This different behavior at large centrality is a result of the even harmonics mainly being driven by the geometric shape of the overlap area, whereas the odd harmonics are mainly driven by fluctuations, which get enhanced through the added nuclear substructure. Since a higher number of binary collisions tends to wash out the details of the sub-nuclear structure, the difference for different $m$ appears at different centralities when considering a large system, such as a PbPb collision, compared to smaller systems, such as OO or NeNe collisions. These differences arise at around the same value of $N_{coll}$ in all systems, therefore appearing at smaller centralities for the smaller systems. To be able to study the effects of different $m$ we therefore will focus on OO collisions when examining the harmonic eccentricities further.

\begin{figure*}[h]
\includegraphics[width=1.\linewidth]{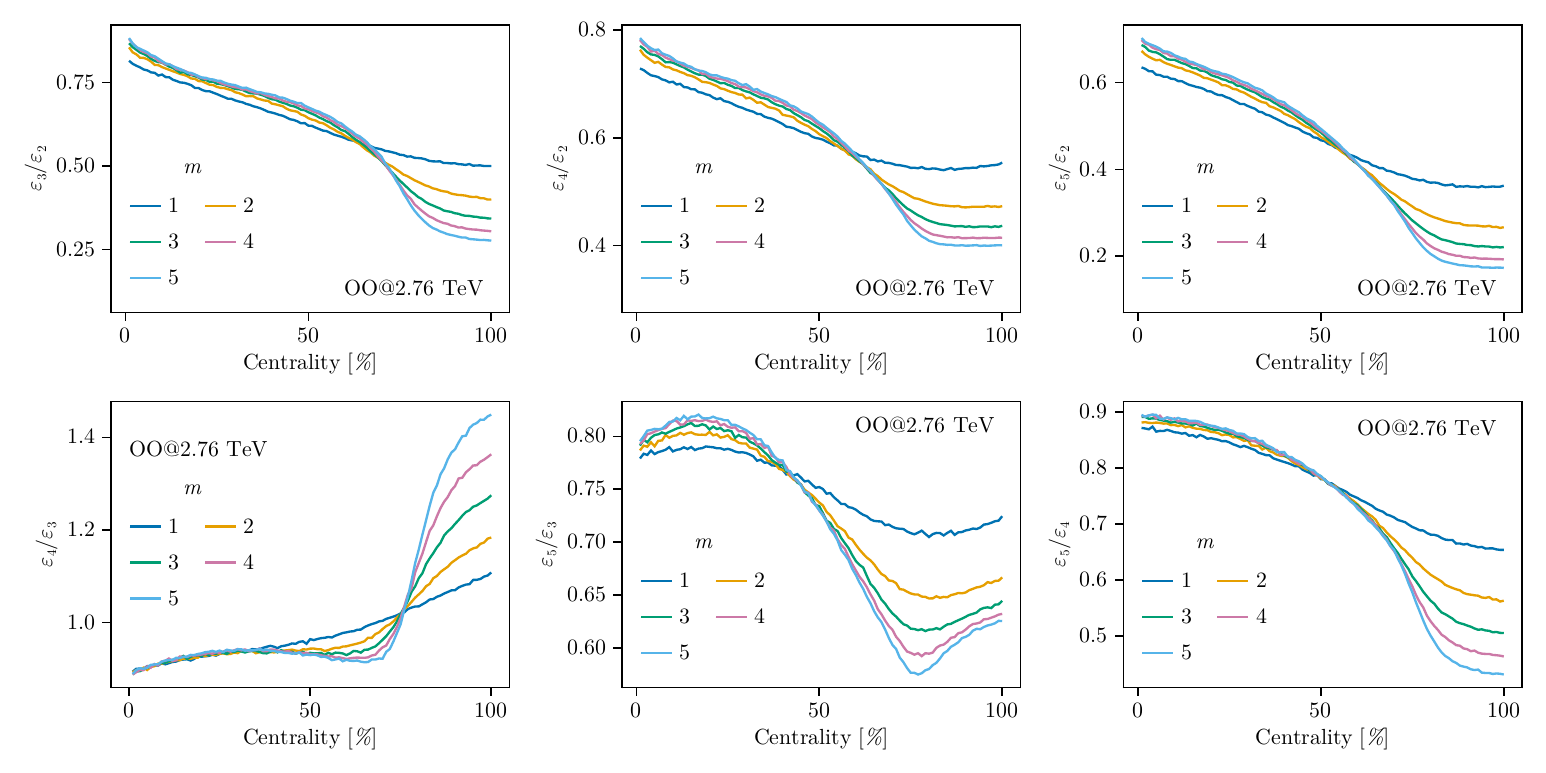}
\caption{Ratio between the initial state eccentricities for $n=2,3,4,5$ as function of centrality for different numbers of partons. All ratios are dominated by the even eccentricities, except $\varepsilon_5/\varepsilon_3$, which only contains odd harmonics, therefore being more sensitive to the value of $m$.}
\label{fig:EccDiffOO}
\end{figure*}

In order to enhance effects appearing due to differences in $m$ and suppressing common effects, we consider the ratios between the eccentricities for OO collisions, as depicted in Fig.~\ref{fig:EccDiffOO}\footnote{The differences between the various harmonics display a similar behavior.}. All ratios, including even eccentricities, are again mainly driven by the geometric shape of the collision and therefore only show differences for different $m$ at large centralities. However, the ratio of the odd eccentricities is driven by fluctuations and therefore displays differences already around $50\%$ centrality.

The ratio in odd harmonics shows additional features, which could help in constraining the number of partons: While being a relatively flat curve for small values of $m$, the curve becomes steeper for larger values of $m$, with its maximum and minimum moving apart. Therefore, the smaller ratios at lower centralities could be leveraged together with the slope when going to larger centralities to constrain the value of $m$ or its distribution. This behavior can be attributed to $\varepsilon_3$ and $\varepsilon_5$ being relatively small for the central collisions, with the triangularity becoming more pronounced at higher centralities due to the sharper profiles at large values of $m$.

\begin{figure}[h]
\includegraphics[width=1.\linewidth]{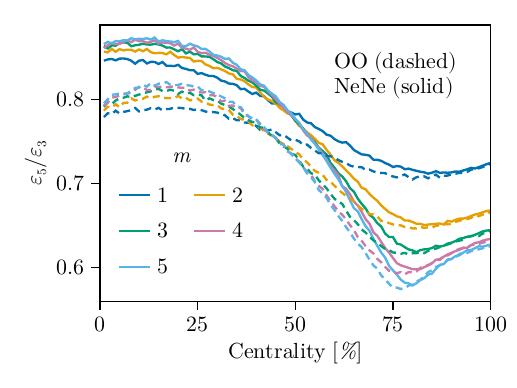}
\caption{Ratio in harmonic eccentricities $\varepsilon_3/\varepsilon_5$ for OO and NeNe collision system. Since the odd eccentricities are dominated by fluctuations instead of the geometric shape of the overlap region, this observable displays sensitivity to $m$ and its distribution at all centralities.}
\label{fig:OONeNe}
\end{figure}

This feature is present in both OO collisions and NeNe collisions (see Fig. \ref{fig:OONeNe}) and therefore could be an interesting addition to the analysis of the current OO and NeNe runs at LHC and possible future light ion runs.

While similar features also appear in other combinations of the eccentricities, such as their ratios, it is sufficient to include one additional observable containing the first two odd harmonics. To reduce uncertainties on the experimental side, we propose the above discussed ration between $\varepsilon_3$ and $\varepsilon_5$. Here, it should be noted that the fifth flow harmonic is not directly proportional to $\varepsilon_5$ anymore, due to non-linear flow effects. Therefore, the observed signal in the flow coefficients might not be as strong in the eccentricities.

\section{Conclusion}

In this work, we proposed a novel way of obtaining the number of partons in each proton and neutron participating in a heavy-ion collision. This method is based on the information obtained, among other through DIS experiments in parton distribution functions and therefore encodes the underlying QCD dynamics governing the structure of the proton. We examined the resulting distribution as a function of momentum transfer and discussed possible implications for the collision, while ensuring a finite number of partons for each sampled proton and neutron. Additionally, we propose possible modifications of this sampling algorithm accounting for different virtualities of the sea-quark and their antiparticles. As final part of the discussion of the sampling method, we sketch the possibility of obtaining a non-zero fluid velocity profile using additional information encoded in transfer momentum distribution functions. 

In the second part of this work, we examined the dependence of different observables on the number of partons $m$ while keeping $m$ fixed. This is achieved by using the initial state estimators for particle number, average transverse momentum, and flow coefficients, namely entropy, energy, and eccentricities. We demonstrate that most observables show little to no sensitivity to the number of constituents, as expected through the inherent averaging dynamics of a heavy-ion collision. However, we were able to demonstrate that the odd eccentricities show some sensitivity to the value of $m$ at small centralities and a comparatively large sensitivity at large centralities. This sensitivity can be increased by considering their ratio, such that a clear change for different values of $m$ can be established already at around $50\%$ centrality.

We propose the inclusion of the ratio of the odd eccentricities as an additional variable in the upcoming analysis of small collision systems, as well as the addition of a fluctuating parton number based on PDFs in the next generations of initial state models, such as \trento.

\section*{Data availability statement}
The data used for this research is available at \cite{github}.

\section*{Acknowledgement}
The authors want to thank Jyotirmoy Roy and Marston Copeland for useful discussions. This work is supported by the U.S. Department of Energy, Office of Science, Office of Nuclear Physics, grant No. DE-FG02-05ER41367.

\clearpage 
\bibliography{references}

@article{Nijs:2021clz,
    author = "Nijs, Govert and van der Schee, Wilke",
    title = "{Predictions and postdictions for relativistic lead and oxygen collisions with the computational simulation code Trajectum}",
    eprint = "2110.13153",
    archivePrefix = "arXiv",
    primaryClass = "nucl-th",
    reportNumber = "CERN-TH-2021-160, MIT-CTP/5333",
    doi = "10.1103/PhysRevC.106.044903",
    journal = "Phys. Rev. C",
    volume = "106",
    number = "4",
    pages = "044903",
    year = "2022"
}

@article{Nijs:2020ors,
    author = {Nijs, Govert and van der Schee, Wilke and G\"ursoy, Umut and Snellings, Raimond},
    title = "{Transverse Momentum Differential Global Analysis of Heavy-Ion Collisions}",
    eprint = "2010.15130",
    archivePrefix = "arXiv",
    primaryClass = "nucl-th",
    reportNumber = "CERN-TH-2020-174, MIT-CTP/5250",
    doi = "10.1103/PhysRevLett.126.202301",
    journal = "Phys. Rev. Lett.",
    volume = "126",
    number = "20",
    pages = "202301",
    year = "2021"
}

@article{Nijs:2020roc,
    author = {Nijs, Govert and van der Schee, Wilke and G\"ursoy, Umut and Snellings, Raimond},
    title = "{Bayesian analysis of heavy ion collisions with the heavy ion computational framework Trajectum}",
    eprint = "2010.15134",
    archivePrefix = "arXiv",
    primaryClass = "nucl-th",
    reportNumber = "CERN-TH-2020-175, MIT-CTP/5251",
    doi = "10.1103/PhysRevC.103.054909",
    journal = "Phys. Rev. C",
    volume = "103",
    number = "5",
    pages = "054909",
    year = "2021"
}

@article{ALICE:2022wpn,
    author = "Acharya, Shreyasi and others",
    collaboration = "ALICE",
    title = "{The ALICE experiment: a journey through QCD}",
    eprint = "2211.04384",
    archivePrefix = "arXiv",
    primaryClass = "nucl-ex",
    reportNumber = "CERN-EP-2022-227",
    doi = "10.1140/epjc/s10052-024-12935-y",
    journal = "Eur. Phys. J. C",
    volume = "84",
    number = "8",
    pages = "813",
    year = "2024"
}

@article{Schenke:2012wb,
    author = "Schenke, Bjoern and Tribedy, Prithwish and Venugopalan, Raju",
    title = "{Fluctuating Glasma initial conditions and flow in heavy ion collisions}",
    eprint = "1202.6646",
    archivePrefix = "arXiv",
    primaryClass = "nucl-th",
    doi = "10.1103/PhysRevLett.108.252301",
    journal = "Phys. Rev. Lett.",
    volume = "108",
    pages = "252301",
    year = "2012"
}

@article{Niemi:2015qia,
    author = "Niemi, H. and Eskola, K. J. and Paatelainen, R.",
    title = "{Event-by-event fluctuations in a perturbative QCD + saturation + hydrodynamics model: Determining QCD matter shear viscosity in ultrarelativistic heavy-ion collisions}",
    eprint = "1505.02677",
    archivePrefix = "arXiv",
    primaryClass = "hep-ph",
    doi = "10.1103/PhysRevC.93.024907",
    journal = "Phys. Rev. C",
    volume = "93",
    number = "2",
    pages = "024907",
    year = "2016"
}

@article{Kurkela:2018vqr,
    author = {Kurkela, Aleksi and Mazeliauskas, Aleksas and Paquet, Jean-Fran{\c{c}}ois and Schlichting, S{\"o}ren and Teaney, Derek},
    title = "{Effective kinetic description of event-by-event pre-equilibrium dynamics in high-energy heavy-ion collisions}",
    eprint = "1805.00961",
    archivePrefix = "arXiv",
    primaryClass = "hep-ph",
    doi = "10.1103/PhysRevC.99.034910",
    journal = "Phys. Rev. C",
    volume = "99",
    number = "3",
    pages = "034910",
    year = "2019"
}

@article{Loizides:2017ack,
    author = "Loizides, Constantin and Kamin, Jason and d'Enterria, David",
    title = "{Improved Monte Carlo Glauber predictions at present and future nuclear colliders}",
    eprint = "1710.07098",
    archivePrefix = "arXiv",
    primaryClass = "nucl-ex",
    doi = "10.1103/PhysRevC.97.054910",
    journal = "Phys. Rev. C",
    volume = "97",
    number = "5",
    pages = "054910",
    year = "2018",
    note = "[Erratum: Phys.Rev.C 99, 019901 (2019)]"
}

@article{Miller:2007ri,
    author = "Miller, Michael L. and Reygers, Klaus and Sanders, Stephen J. and Steinberg, Peter",
    title = "{Glauber modeling in high energy nuclear collisions}",
    eprint = "nucl-ex/0701025",
    archivePrefix = "arXiv",
    doi = "10.1146/annurev.nucl.57.090506.123020",
    journal = "Ann. Rev. Nucl. Part. Sci.",
    volume = "57",
    pages = "205--243",
    year = "2007"
}

@article{Moreland:2014oya,
    author = "Moreland, J. Scott and Bernhard, Jonah E. and Bass, Steffen A.",
    title = "{Alternative ansatz to wounded nucleon and binary collision scaling in high-energy nuclear collisions}",
    eprint = "1412.4708",
    archivePrefix = "arXiv",
    primaryClass = "nucl-th",
    doi = "10.1103/PhysRevC.92.011901",
    journal = "Phys. Rev. C",
    volume = "92",
    number = "1",
    pages = "011901",
    year = "2015"
}

@article{Ke:2016jrd,
    author = "Ke, Weiyao and Moreland, J. Scott and Bernhard, Jonah E. and Bass, Steffen A.",
    title = "{Constraints on rapidity-dependent initial conditions from charged particle pseudorapidity densities and two-particle correlations}",
    eprint = "1610.08490",
    archivePrefix = "arXiv",
    primaryClass = "nucl-th",
    doi = "10.1103/PhysRevC.96.044912",
    journal = "Phys. Rev. C",
    volume = "96",
    number = "4",
    pages = "044912",
    year = "2017"
}

@article{Moreland:2018gsh,
    author = "Moreland, J. Scott and Bernhard, Jonah E. and Bass, Steffen A.",
    title = "{Bayesian calibration of a hybrid nuclear collision model using p-Pb and Pb-Pb data at energies available at the CERN Large Hadron Collider}",
    eprint = "1808.02106",
    archivePrefix = "arXiv",
    primaryClass = "nucl-th",
    doi = "10.1103/PhysRevC.101.024911",
    journal = "Phys. Rev. C",
    volume = "101",
    number = "2",
    pages = "024911",
    year = "2020"
}

@article{Soeder:2023vdn,
    author = "Soeder, Derek and Ke, Weiyao and Paquet, J. -F. and Bass, Steffen A.",
    title = "{Bayesian parameter estimation with a new three-dimensional initial-conditions model for ultrarelativistic heavy-ion collisions}",
    eprint = "2306.08665",
    archivePrefix = "arXiv",
    primaryClass = "nucl-th",
    month = "6",
    year = "2023"
}

@article{STAR:2021mii,
    author = "Abdallah, Mohamed and others",
    collaboration = "STAR",
    title = "{Search for the chiral magnetic effect with isobar collisions at $\sqrt {s_{NN}}$=200 GeV by the STAR Collaboration at the BNL Relativistic Heavy Ion Collider}",
    eprint = "2109.00131",
    archivePrefix = "arXiv",
    primaryClass = "nucl-ex",
    doi = "10.1103/PhysRevC.105.014901",
    journal = "Phys. Rev. C",
    volume = "105",
    number = "1",
    pages = "014901",
    year = "2022"
}

@article{Giacalone:2025vxa,
    author = "Giacalone, Giuliano and others",
    title = "{Nuclear Physics Confronts Relativistic Collisions Of Isobars}",
    eprint = "2507.01454",
    archivePrefix = "arXiv",
    primaryClass = "nucl-ex",
    month = "7",
    year = "2025"
}

@article{Harland-Lang:2014zoa,
    author = "Harland-Lang, L. A. and Martin, A. D. and Motylinski, P. and Thorne, R. S.",
    title = "{Parton distributions in the LHC era: MMHT 2014 PDFs}",
    eprint = "1412.3989",
    archivePrefix = "arXiv",
    primaryClass = "hep-ph",
    reportNumber = "LCTS-2014-47, IPPP-14-97, DCPT-14-194",
    doi = "10.1140/epjc/s10052-015-3397-6",
    journal = "Eur. Phys. J. C",
    volume = "75",
    number = "5",
    pages = "204",
    year = "2015"
}

@article{Gribov:1972ri,
    author = "Gribov, V. N. and Lipatov, L. N.",
    title = "{Deep inelastic e p scattering in perturbation theory}",
    reportNumber = "IPTI-381-71",
    journal = "Sov. J. Nucl. Phys.",
    volume = "15",
    pages = "438--450",
    year = "1972"
}

@article{Dokshitzer:1977sg,
    author = "Dokshitzer, Yuri L.",
    title = "{Calculation of the Structure Functions for Deep Inelastic Scattering and e+ e- Annihilation by Perturbation Theory in Quantum Chromodynamics.}",
    journal = "Sov. Phys. JETP",
    volume = "46",
    pages = "641--653",
    year = "1977"
}

@article{Altarelli:1977zs,
    author = "Altarelli, Guido and Parisi, G.",
    title = "{Asymptotic Freedom in Parton Language}",
    reportNumber = "LPTENS-77-6",
    doi = "10.1016/0550-3213(77)90384-4",
    journal = "Nucl. Phys. B",
    volume = "126",
    pages = "298--318",
    year = "1977"
}

@article{Bernhard:2016tnd,
    author = "Bernhard, Jonah E. and Moreland, J. Scott and Bass, Steffen A. and Liu, Jia and Heinz, Ulrich",
    title = "{Applying Bayesian parameter estimation to relativistic heavy-ion collisions: simultaneous characterization of the initial state and quark-gluon plasma medium}",
    eprint = "1605.03954",
    archivePrefix = "arXiv",
    primaryClass = "nucl-th",
    doi = "10.1103/PhysRevC.94.024907",
    journal = "Phys. Rev. C",
    volume = "94",
    number = "2",
    pages = "024907",
    year = "2016"
}

@article{Bernhard:2019bmu,
    author = "Bernhard, Jonah E. and Moreland, J. Scott and Bass, Steffen A.",
    title = "{Bayesian estimation of the specific shear and bulk viscosity of quark{\textendash}gluon plasma}",
    doi = "10.1038/s41567-019-0611-8",
    journal = "Nature Phys.",
    volume = "15",
    number = "11",
    pages = "1113--1117",
    year = "2019"
}

@article{JETSCAPE:2020mzn,
    author = "Everett, D. and others",
    collaboration = "JETSCAPE",
    title = "{Multisystem Bayesian constraints on the transport coefficients of QCD matter}",
    eprint = "2011.01430",
    archivePrefix = "arXiv",
    primaryClass = "hep-ph",
    doi = "10.1103/PhysRevC.103.054904",
    journal = "Phys. Rev. C",
    volume = "103",
    number = "5",
    pages = "054904",
    year = "2021"
}

@article{Harris:1996zx,
    author = "Harris, John W. and Muller, Berndt",
    title = "{The Search for the quark - gluon plasma}",
    eprint = "hep-ph/9602235",
    archivePrefix = "arXiv",
    reportNumber = "DUKE-TH-96-105",
    doi = "10.1146/annurev.nucl.46.1.71",
    journal = "Ann. Rev. Nucl. Part. Sci.",
    volume = "46",
    pages = "71--107",
    year = "1996"
}

@article{Harris:2023tti,
    author = {Harris, John W. and M{\"u}ller, Berndt},
    title = "{''QGP Signatures'' Revisited}",
    eprint = "2308.05743",
    archivePrefix = "arXiv",
    primaryClass = "hep-ph",
    doi = "10.1140/epjc/s10052-024-12533-y",
    journal = "Eur. Phys. J. C",
    volume = "84",
    number = "3",
    pages = "247",
    year = "2024"
}

@article{PHENIX:2003iij,
    author = "Adler, S. S. and others",
    collaboration = "PHENIX",
    title = "{Identified charged particle spectra and yields in Au+Au collisions at S(NN)**1/2 = 200-GeV}",
    eprint = "nucl-ex/0307022",
    archivePrefix = "arXiv",
    doi = "10.1103/PhysRevC.69.034909",
    journal = "Phys. Rev. C",
    volume = "69",
    pages = "034909",
    year = "2004"
}

@article{PHENIX:2003qra,
    author = "Adler, S. S. and others",
    collaboration = "PHENIX",
    title = "{Elliptic flow of identified hadrons in Au+Au collisions at s(NN)**(1/2) = 200-GeV}",
    eprint = "nucl-ex/0305013",
    archivePrefix = "arXiv",
    doi = "10.1103/PhysRevLett.91.182301",
    journal = "Phys. Rev. Lett.",
    volume = "91",
    pages = "182301",
    year = "2003"
}

@article{STAR:2008med,
    author = "Abelev, B. I. and others",
    collaboration = "STAR",
    title = "{Systematic Measurements of Identified Particle Spectra in $p p, d^+$ Au and Au+Au Collisions from STAR}",
    eprint = "0808.2041",
    archivePrefix = "arXiv",
    primaryClass = "nucl-ex",
    doi = "10.1103/PhysRevC.79.034909",
    journal = "Phys. Rev. C",
    volume = "79",
    pages = "034909",
    year = "2009"
}

@article{ALICE:2013mez,
    author = "Abelev, Betty and others",
    collaboration = "ALICE",
    title = "{Centrality dependence of $\pi$, K, p production in Pb-Pb collisions at $\sqrt{s_{NN}}$ = 2.76 TeV}",
    eprint = "1303.0737",
    archivePrefix = "arXiv",
    primaryClass = "hep-ex",
    reportNumber = "CERN-PH-EP-2013-019",
    doi = "10.1103/PhysRevC.88.044910",
    journal = "Phys. Rev. C",
    volume = "88",
    pages = "044910",
    year = "2013"
}

@article{HotQCD:2014kol,
    author = "Bazavov, A. and others",
    collaboration = "HotQCD",
    title = "{Equation of state in ( 2+1 )-flavor QCD}",
    eprint = "1407.6387",
    archivePrefix = "arXiv",
    primaryClass = "hep-lat",
    reportNumber = "BNL-105928-2014-JA",
    doi = "10.1103/PhysRevD.90.094503",
    journal = "Phys. Rev. D",
    volume = "90",
    pages = "094503",
    year = "2014"
}

@article{Christiansen:2014ypa,
    author = "Christiansen, Nicolai and Haas, Michael and Pawlowski, Jan M. and Strodthoff, Nils",
    title = "{Transport Coefficients in Yang--Mills Theory and QCD}",
    eprint = "1411.7986",
    archivePrefix = "arXiv",
    primaryClass = "hep-ph",
    doi = "10.1103/PhysRevLett.115.112002",
    journal = "Phys. Rev. Lett.",
    volume = "115",
    number = "11",
    pages = "112002",
    year = "2015"
}

@article{Policastro:2001yc,
    author = "Policastro, G. and Son, Dan T. and Starinets, Andrei O.",
    title = "{The Shear viscosity of strongly coupled N=4 supersymmetric Yang-Mills plasma}",
    eprint = "hep-th/0104066",
    archivePrefix = "arXiv",
    reportNumber = "NYU-TH-01-04-02, SNS-PH-01-05",
    doi = "10.1103/PhysRevLett.87.081601",
    journal = "Phys. Rev. Lett.",
    volume = "87",
    pages = "081601",
    year = "2001"
}

@article{Giacalone:2024luz,
    author = "Giacalone, Giuliano and others",
    title = "{Exploiting Ne20 Isotopes for Precision Characterizations of Collectivity in Small Systems}",
    eprint = "2402.05995",
    archivePrefix = "arXiv",
    primaryClass = "nucl-th",
    reportNumber = "CERN-TH-2024-021",
    doi = "10.1103/k8rb-jgvq",
    journal = "Phys. Rev. Lett.",
    volume = "135",
    number = "1",
    pages = "012302",
    year = "2025"
}

@article{Giacalone:2021clp,
    author = {Giacalone, Giuliano and Schenke, Bj{\"o}rn and Shen, Chun},
    title = "{Constraining the Nucleon Size with Relativistic Nuclear Collisions}",
    eprint = "2111.02908",
    archivePrefix = "arXiv",
    primaryClass = "nucl-th",
    doi = "10.1103/PhysRevLett.128.042301",
    journal = "Phys. Rev. Lett.",
    volume = "128",
    number = "4",
    pages = "042301",
    year = "2022"
}

@article{Shen:2017bsr,
    author = {Shen, Chun and Schenke, Bj{\"o}rn},
    title = "{Dynamical initial state model for relativistic heavy-ion collisions}",
    eprint = "1710.00881",
    archivePrefix = "arXiv",
    primaryClass = "nucl-th",
    doi = "10.1103/PhysRevC.97.024907",
    journal = "Phys. Rev. C",
    volume = "97",
    number = "2",
    pages = "024907",
    year = "2018"
}

@article{Lai:1999wy,
    author = "Lai, H. L. and Huston, J. and Kuhlmann, S. and Morfin, J. and Olness, Fredrick I. and Owens, J. F. and Pumplin, J. and Tung, W. K.",
    collaboration = "CTEQ",
    title = "{Global QCD analysis of parton structure of the nucleon: CTEQ5 parton distributions}",
    eprint = "hep-ph/9903282",
    archivePrefix = "arXiv",
    reportNumber = "MSUHEP-903100, FERMILAB-PUB-00-266-E",
    doi = "10.1007/s100529900196",
    journal = "Eur. Phys. J. C",
    volume = "12",
    pages = "375--392",
    year = "2000"
}

@article{Kovchegov:2014kua,
    author = "Kovchegov, Yuri V.",
    editor = "Praszalowicz, Michal",
    title = "{Brief Review of Saturation Physics}",
    eprint = "1410.7722",
    archivePrefix = "arXiv",
    primaryClass = "hep-ph",
    doi = "10.5506/APhysPolB.45.2241",
    journal = "Acta Phys. Polon. B",
    volume = "45",
    number = "12",
    pages = "2241--2256",
    year = "2014"
}

@phdthesis{Giacalone:2020ymy,
    author = "Giacalone, Giuliano",
    title = "{A matter of shape: seeing the deformation of atomic nuclei at high-energy colliders}",
    eprint = "2101.00168",
    archivePrefix = "arXiv",
    primaryClass = "nucl-th",
    reportNumber = "tel-03185076, 2020UPASP072",
    school = "U. Paris-Saclay",
    year = "2020"
}

@article{Xu:2021uar,
    author = "Xu, Hao-jie and Zhao, Wenbin and Li, Hanlin and Zhou, Ying and Chen, Lie-Wen and Wang, Fuqiang",
    title = "{Probing nuclear structure with mean transverse momentum in relativistic isobar collisions}",
    eprint = "2111.14812",
    archivePrefix = "arXiv",
    primaryClass = "nucl-th",
    doi = "10.1103/PhysRevC.108.L011902",
    journal = "Phys. Rev. C",
    volume = "108",
    number = "1",
    pages = "L011902",
    year = "2023"
}

@article{Jia:2021oyt,
    author = "Jia, Jiangyong and Zhang, Chunjian",
    title = "{Scaling approach to nuclear structure in high-energy heavy-ion collisions}",
    eprint = "2111.15559",
    archivePrefix = "arXiv",
    primaryClass = "nucl-th",
    doi = "10.1103/PhysRevC.107.L021901",
    journal = "Phys. Rev. C",
    volume = "107",
    number = "2",
    pages = "L021901",
    year = "2023"
}

@article{Nijs:2021kvn,
    author = "Nijs, Govert and van der Schee, Wilke",
    title = "{Inferring nuclear structure from heavy isobar collisions using Trajectum}",
    eprint = "2112.13771",
    archivePrefix = "arXiv",
    primaryClass = "nucl-th",
    reportNumber = "CERN-TH-2021-229, MIT-CTP/5383",
    doi = "10.21468/SciPostPhys.15.2.041",
    journal = "SciPost Phys.",
    volume = "15",
    number = "2",
    pages = "041",
    year = "2023"
}

@article{Li:2018oec,
    author = "Li, Hanlin and Xu, Hao-jie and Zhao, Jie and Lin, Zi-Wei and Zhang, Hanzhong and Wang, Xiaobao and Shen, Caiwan and Wang, Fuqiang",
    title = "{Multiphase transport model predictions of isobaric collisions with nuclear structure from density functional theory}",
    eprint = "1808.06711",
    archivePrefix = "arXiv",
    primaryClass = "nucl-th",
    doi = "10.1103/PhysRevC.98.054907",
    journal = "Phys. Rev. C",
    volume = "98",
    number = "5",
    pages = "054907",
    year = "2018"
}

@article{Xu:2021vpn,
    author = "Xu, Hao-jie and Li, Hanlin and Wang, Xiaobao and Shen, Caiwan and Wang, Fuqiang",
    title = "{Determine the neutron skin type by relativistic isobaric collisions}",
    eprint = "2103.05595",
    archivePrefix = "arXiv",
    primaryClass = "nucl-th",
    doi = "10.1016/j.physletb.2021.136453",
    journal = "Phys. Lett. B",
    volume = "819",
    pages = "136453",
    year = "2021"
}

@article{Huang:2025cjm,
    author = "Huang, Shengli and Jia, Jiangyong and Zhang, Chunjian",
    title = "{Symmetric-asymmetric collision comparison: Disentangling nuclear structure and subnucleonic structure effects for small system flow}",
    eprint = "2507.16162",
    archivePrefix = "arXiv",
    primaryClass = "nucl-th",
    doi = "10.1016/j.physletb.2025.139926",
    journal = "Phys. Lett. B",
    volume = "870",
    pages = "139926",
    year = "2025"
}

@misc{github,
  author = {Andreas Kirchner, Steffen A. Bass},
  title = {{Fluctuating-constituents}},
  howpublished = {\url{https://github.com/AndreasKirchner/Fluctuating-constituents}},
  year = {Year of last update/access},
  note = {Accessed: 2025-11-12}
}
\clearpage

\end{document}